\documentclass[epjST]{svjour}
\usepackage{graphics}
\usepackage{textcomp}
\usepackage{expl3}
\usepackage{gensymb}
\usepackage[mathscr]{euscript}
\usepackage{bm}
\usepackage{breqn}
\begin{document}
\title{Sweeping by Sessile Drop Coalescence}
\author{Jonathan M. Ludwicki\thanks{\email{jml536@cornell.edu}} \and Paul H. Steen\thanks{\email{phs7@cornell.edu}}}
\institute{Robert Frederick Smith School of Chemical and Biomolecular Engineering, Cornell University, Ithaca, NY, 14853, USA}
\abstract{During coalescence of liquid drops contacting a solid, the liquid sweeps wetted and solid-projected areas. The extent of sweeping dictates the performance of devices such as self-cleaning surfaces, anti-frost coatings, water harvesters, and dropwise condensers. For these applications, weakly- and non-wetting solid substrates are preferred as they enhance drop dynamical behavior. Accordingly, our coalescence studies here are restricted to drops with contact angle $90 \degree \le \theta_{0} \le 180 \degree$. Binary sessile drop coalescence is the focus, with volume of fluid simulations employed as the primary tool. The simulations, which incorporate a Kistler dynamic contact angle model, are first validated against three different experimental substrate systems and then used to study the influence of solid wettability on sweeping by modifying $\theta_{0}$. With increasing $\theta_{0}$ up to 150$\degree$, wetted and projected swept areas both increase as drop center of mass heightens. For $\theta_{0} \ge 150\degree$, coalescence-induced drop jumping occurs owing to the decreasing wettability of the substrate and a focusing of liquid momentum due to the symmetry-breaking solid. In this regime, projected swept area continues to increase with $\theta_0$ while wetted swept area reaches a maximum and then decreases. The sweeping results are interpreted using the mechanical energy balance from hydrodynamic theory and also compared to free drop coalescence.} 

\maketitle
\section{Introduction}
\begin{figure}[t]
\centering
\includegraphics{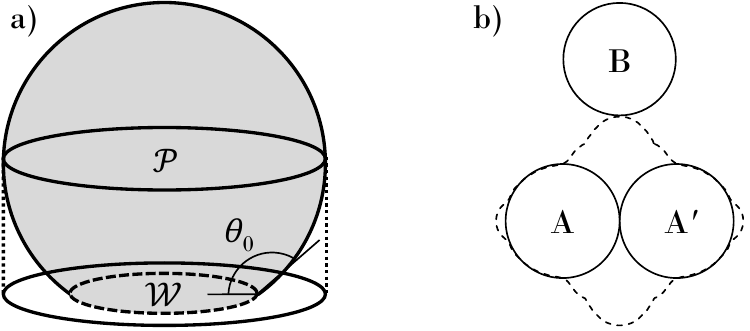}
\caption[]{a) Sessile drop: initial contact angle $\theta_0 > 90\degree$, projected $\mathscr{P}$ and wetted $\mathscr{W}$ contact areas. b) Projected view (top) of coalescing drops $A-A'$ with cumulative swept area (dashed) just reaching drop $B$, perhaps initiating another coalescence.}
\label{fig:1.1}
\end{figure}

\begin{figure}[t]
\centering
\resizebox{1.0\hsize}{!}{\includegraphics*{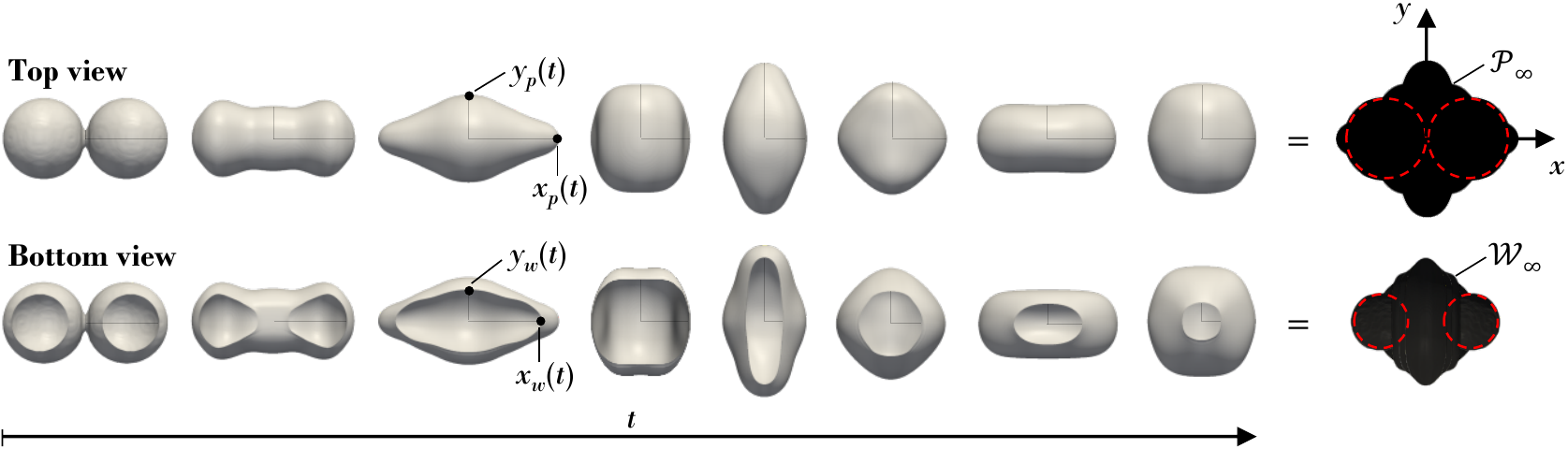}}
\caption[]{Instantaneous projected $\mathscr{P}(t)$ (top row) and wetted $\mathscr{W}(t)$ (bottom row) swept areas, with cumulative projected $\mathscr{P}_\infty$ (top right) and wetted $\mathscr{W}_\infty$ (bottom right) swept areas. Circles on the silhouettes (red dashed) represent the initial areas.}
\label{fig:1.2}
\end{figure}

For coalescing capillary drops, liquid advances and recedes along a solid surface evolving complicated patterns of wetted and projected areas. Instantaneous wetted $\mathscr{W}$ is the area on the surface enclosed by the contact line, while instantaneous projected $\mathscr{P}$ is the area of an imaginary shadow cast from above by parallel projection of light onto the surface. Projected exceeds wetted area for drops on non-wetting and hydrophobic surfaces where initial contact angle $\theta_{0} > 90 \degree$, Figure \ref{fig:1.1}a. There is recent interest in the behaviors of drops on surfaces where wetted and projected areas differ\,\cite{Takeda2002,Tsai2009,Heydari2013,Dash2014,Tavakoli2015}, including sessile drop coalescence. During sessile drop coalescence, two or more drops in contact with a surface merge to form a final drop. This is a dynamic event where the liquid exhibits a complex shape evolution while progressing towards a final equilibrium configuration \cite{Menchaca-Rocha2001,Nilsson2011,Kavehpour2015}. In some instances, wetted area must change topology from initially disconnected disks to a connected disk, for example, Figure \ref{fig:1.2}. As drops merge into a single body of fluid, liquid sweeping accumulates both a cumulative wetted $\mathscr{W}_\infty$ and cumulative projected $\mathscr{P}_\infty$ area.
Wetted sweeping occurs due to contact line motion, while projected sweeping results from dynamics of the bulk liquid above the surface. $\mathscr{P}_\infty = \mathscr{W}_\infty$ on wetting and hydrophilic surfaces, where $\theta_0 < 90 \degree$. However, $\mathscr{P}_\infty > \mathscr{W}_\infty$ for sessile drop coalescence on non-wetting and hydrophobic surfaces where $\theta_0 > 90 \degree$ since, for these, the body of the liquid overhangs the contact line during the coalescence event. Depending on the interplay of solid, liquid, and gas phases, the area swept during a coalescence event can vary dramatically, reflecting how interfacial energy is partitioned into kinetic energy and dissipation.

Understanding swept areas is important both fundamentally and from an applications perspective. For example, self-cleaning surfaces leverage sweeping by sessile drop coalescence to collect contaminants and other foreign debris for removal \cite{Wisdom2013}. Anti-frost coatings rely on coalescence to eject drops before freezing can occur, with multiple drops merging via sweeping during the removal process \cite{Boreyko2013}. The growth rate of drops on water harvesting devices is strongly accelerated by multi-drop coalescence events, which leads to enhanced liquid collection rates \cite{Beysens2006}. Finally, in dropwise condensation heat transfer processes, a principal mechanism of fresh surface generation for renucleation is the sweeping up of drops by a coalescence event \cite{Meakin1992,Macner2014}. These swept regions have a significant impact on heat transfer as smaller, renucleated drops have a higher heat flux as compared to larger, more insulating drops \cite{Kim2011}. A sketch of multi-drop coalescence sweeping is shown in Figure \ref{fig:1.1}b.

A typical coalescence sequence is demonstrated in Figure \ref{fig:1.2}. At the initiation of coalescence between two sessile drops, a small liquid neck forms with large curvature. This curvature generates a significant capillary force in the neck which drives a rapid expansion of the liquid bridge \cite{Eggers1999,Duchemin2003}. As the bridge expands, capillary waves travel outward along the $x$-axis towards the ends of the drops. Once the capillary waves reach the ends, they are reflected back towards the coalescence initiation point. This causes the liquid to contract along the $x$-axis and drives a continued expansion in the $y$-direction until a maximum is reached. The liquid then exhibits damped oscillations that alternate between the $x$- and $y$-directions, with the contact line eventually pinning or, perhaps, with the liquid jumping off the surface. After contact line pinning, remaining kinetic energy is dissipated solely by bulk viscosity and the liquid settles to a final equilibrium shape. For high surface tension liquids like water or mercury, the coalescence event is a strongly underdamped oscillation where inertia and interfacial energy compete while being mitigated by contact line and bulk viscous dissipation. The contact line contribution can be significant, as shown later.

An early formal study on sessile drop coalescence was published in 2001 by Menchaca-Rocha and collaborators \cite{Menchaca-Rocha2001}. In this work, experiments and numerical simulations were employed to elucidate the shape evolution of two identical mercury drops coalescing on a super-non-wettable surface. Since then, numerous papers on sessile drop coalescence have appeared. Topics of investigation have included power law growth of the liquid bridge after coalescence initiation \cite{Ristenpart2006,Sellier2009,Lee2012,Hernandez-Sanchez2012,Sui2013,Eddi2013,Mitra2015}, overall shape evolution \cite{Nilsson2011,Gokhale2004,Kapur2007,Liao2008,Lai2010,Wang2010,Zhu2017,Somwanshi2018}, and long-time contact line relaxation dynamics \cite{Andrieu2002,Narhe2004,Narhe2005,Beysens2006a,Narhe2008}. Following the 2009 discovery of self-propelled dropwise condensate removal on superhydrophobic surfaces \cite{Narhe2009,Boreyko2009}, a majority of these studies have explored coalescence-induced out-of-plane drop jumping \cite{Boreyko2010,Wang2011,Peng2013,Nam2013a,Liu2014,Liu2014a,Enright2014,Liu2015a,Nam2015,Farokhirad2015,Wang2016,Cha2016,Attarzadeh2017,Wang2017,Sheng2017,Mouterde2017,Wang2017a,Wasserfall2017,Zhang2017,Shi2018,Chu2018,Gao2018,Chen2018,Gao2018a,Yuan2019a}. For their utility in interpreting experimental observations, computational fluid dynamics simulations of sessile drop coalescence have also become popular over the last decade \cite{Wang2010b,Ahmadlouydarab2014,Moghtadernejad2015,Seo2017}. The coalescence sweeping phenomenon has not been studied previously, as far as we are aware.

In this paper, numerical simulations are performed to investigate sweeping by binary sessile drop coalescence for identical drops having no approach velocity in a solid-water-air system. First, the numerical model, simulation domain, and computational procedure are described. The model is then validated versus three experimental cases of sessile water drop coalescence. After verifying the model, the influence of the surface wettability on sweeping is explored by changing the initial contact angle of the coalescing drops ($\theta_{0}$). It is noted that testing is limited to cases where $ 90\degree \leq \theta_{0} \leq 180\degree$ since hydrophobic and/or superhydrophobic surfaces are typically employed for self-cleaning, anti-frost, water harvesting, and dropwise condensation applications. Wetted and projected swept areas are then quantified, with an energy accounting performed to help understand the sweeping behavior. The sessile drop coalescence results are further contrasted with the free drop coalescence case to investigate how the surface constraint influences sweeping. Finally, the time duration of sweeping is explored. This study is a first to quantify sweeping by sessile drop coalescence and to address the role of surface wettability in determining the extent of area swept, to our knowledge.

\section{Numerical Method}

\subsection{Simulation Perspective}
Simulating interface dynamics for incompressible two-phase flows is an active research topic \cite{Mirjalili2019}. Adding moving contact lines further complicates the problem. We choose a VOF approach with a diffuse interface based on its success in prior studies with moving contact lines. Alternative approaches to incorporating moving contact lines split between those where the dynamic contact angle is an input \cite{Dodds2012,Huang2019} and those where it is an output \cite{Shikhmurzaev2007,Liu2016} of the numerical simulation. We choose as an `input' contact-line model the Kistler empirical correlation \cite{Kistler1993} since this experimentally determined correlation has demonstrated engineering utility. The smearing of the liquid-gas interface inherent to our VOF model allows for compatibility of Kistler with a no-slip boundary condition imposed on the solid surface. For comparison, a sharp interface model would require some form of stress relief, for example, by way of a slip length. Notably, the Kistler method depends only on the advancing and receding contact angles, two parameters which can be measured independently in experiment. As a consequence, an important and distinguishing feature of our simulation setup is that no parameters are left free for fitting to experiment.

\subsection{Volume of Fluid Formulation}
Sessile drop coalescence is simulated in OpenFOAM, an open source computational fluid dynamics software \cite{Weller1998,OpenFOAM}. A modified version of the interFoam VOF solver is employed where artificial anti-diffusive surface fluxes are removed to allow for a more accurate prediction of the capillary-dominated coalescence flow \cite{Wasserfall2017}. This brings the time response of the simulation dynamics into line with the experimental results, Figure S1, Supplementary Material. A VOF method is chosen for its advantages in mass conservation. The interFoam solver is employed for incompressible, laminar, two-phase fluid flow with each computational cell containing some volume fraction of liquid ($\alpha$) where 0 $\leq$ $\alpha$ $\leq$ 1. In any cell, fluid properties like density ($\rho$) and dynamic viscosity ($\mu$) are a weighted average based on the volume fraction of liquid,
\begin{equation}
  \rho = \alpha\rho_{l}+(1-\alpha)\rho_{g},
\end{equation}
\begin{equation}
  \mu = \alpha\mu_{l}+(1-\alpha)\mu_{g},
\end{equation}
where subscripts $l$ and $g$ denote liquid and gas phases, respectively. Since $\alpha$ is a conserved scalar via the incompressibility condition, it satisfies the advection equation
\begin{equation} \label{eq:advection}
  \frac{\partial\alpha}{\partial t} \, + \bm{\nabla}\cdot(\alpha\bm{u}) = 0.
\end{equation}
Note the lack of an artificial interface compression term in this equation. In the simulations reported below, interface thicknesses vary in time but stay below a maximum of $5\%$ of the computational domain size. In the standard way for a VOF method, continuity and momentum equations govern the flow,
\begin{equation} \label{eq:continuity}
  \bm{\nabla}\cdot \bm{u} = 0,
\end{equation}
\begin{equation} \label{eq:momentum}
  \frac{\partial\rho \bm{u}}{\partial t} \, + \bm{\nabla}\cdot(\rho\bm{uu}) = -(\bm{\nabla}\cdot P)\: + \:\bm{\nabla}\cdot\Big(\mu\,(\bm{\nabla}\bm{u}+\bm{\nabla}\bm{u}^T)\Bigr)+\rho\,\bm{g} + \bm{F_b},
\end{equation}
where $\bm{u} = (u_x, u_y, u_z$) is the velocity, $t$ is time, $P$ is pressure, and $\bm{g}$ is the gravity vector. Here, $\bm{F_b} = \sigma_{lg}\kappa\bm{\nabla}\alpha$ is the surface tension body force vector modeled by the continuum surface force method of Brackbill \cite{Brackbill1992} where $\sigma_{lg}$ is liquid-gas surface tension and $\kappa = -\bm{\nabla}\cdot\Big(\frac{\bm{\nabla}\alpha}{|\bm{\nabla}\alpha|}\Bigr)$ is the liquid-gas interfacial curvature.
By definition, the surface tension body force acts solely on liquid-gas interfacial cells. Equations (\ref{eq:advection}), (\ref{eq:continuity}), and (\ref{eq:momentum}) are solved subject to a moving contact line model, specified next.

\subsection{Dynamic Contact Angle Model}
\begin{figure}[b]
\resizebox{0.87\hsize}{!}{\includegraphics*{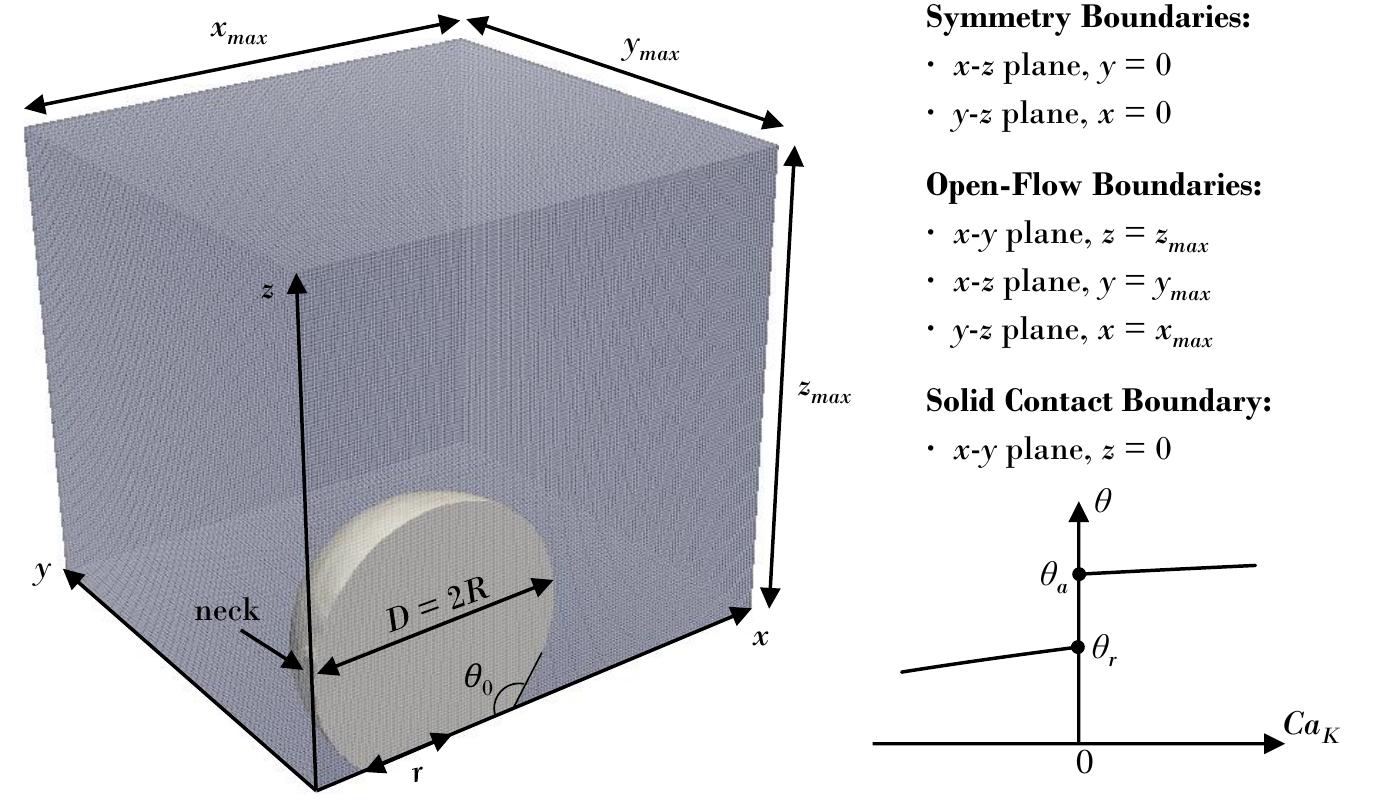}}
\caption[]{Computational domain: flow is bounded by symmetry, open-flow, and solid contact boundaries. At the moving contact line, the Kistler model is implemented, Equations (6), (7) and (8), sketched (lower right) and showing stick for $\theta_{r} \le \theta \le \theta_{a}$ and slip otherwise.}
\label{fig:1.3}
\end{figure}

To account for fluid interactions with the surface, the dynamic contact angle model of Kistler \cite{Kistler1993} is employed for cases of advancing and receding contact line. The Kistler model relates the dynamic contact angle $\theta$ to the velocity of the contact line $U_{CL}$, using in its scaled form the Kistler capillary number $Ca_K = \mu U_{CL}/\sigma_{lg}$. Kistler has been shown to accurately represent the behavior of moving contact lines in other drop-based simulations, including sessile drop coalescence, without any fitting parameters \cite{Moghtadernejad2015,Sikalo2005,Roisman2008,Saha2009,Graham2012,Xu2018}. 
For advancing and receding contact lines, respectively, 
\begin{equation}
  \theta = f[f^{-1}(\theta_a)+Ca_K] \ for \ U_{CL} > 0,
\end{equation}
\begin{equation}
  \theta = f[f^{-1}(\theta_r)-Ca_K] \ for \ U_{CL} < 0,
\end{equation}
where $\theta_a$ is the advancing contact angle, $\theta_r$ is the receding contact angle, and $f$ is the empirical Hoffman function, based on a systematic study of silicone oils displacing air in glass capillaries \cite{Hoffman1975}, expressed mathematically by Kistler as
\begin{equation}
  f(x) = \cos^{-1}\bigg\{1-2\tanh{\bigg[5.16\Big(\frac{x}{1+1.31x^{0.99}}\Big)^{0.706}\bigg]\bigg\}}
\end{equation}
with $0 \leq f \leq \pi$. When $U_{CL} = 0$, the contact line is stationary and the contact angle is free to exhibit values between $\theta_r$ and $\theta_a$. 
Note that $\theta_0$ does not appear explicitly in the model but is bounded by $\theta_r$ $<$ $\theta_0$ $<$ $\theta_a$. See Figure \ref{fig:1.3} (inset) for a sketch of $\theta$ vs. $Ca_K$ response.

\begin{table}[!b]
\centering
\caption[]{Simulation fluid properties: water (subscript $l$) and air (subscript $g$).}
\begin{tabular}{|c|c|c|c|c|}
    \hline
    $\rho_l$ (kg/m\textsuperscript{3}) & $\rho_g$ (kg/m\textsuperscript{3}) & $\mu_l$ (mPa$\,\cdot\,$s) & $\mu_g$ (mPa$\,\cdot\,$s) & $\sigma_{lg}$ (J/m\textsuperscript{2})\\
    \hline
    \hline
    998.0 & 1.204 & 0.998 & 0.0181 & 0.072\\
    \hline
\end{tabular}
\label{tab:1.1}
\end{table}

\subsection{Computational Domain and Procedure}
A 1.4 mm x 1.3 mm x 1.3 mm (\textit{x} by \textit{y} by \textit{z}) computational domain comprised of cubic cells with edge length $\epsilon = 10 \ \mu$m is used to study the influence of surface wettability on coalescence sweeping (Figure \ref{fig:1.3}). In simulating the coalescence of two identical drops with no approach velocity, two symmetry planes are employed to reduce the computational load. At the top and sides of the domain, a fixed pressure boundary with zero velocity gradient is applied. The bottom of the domain is a no-slip solid contact boundary. Near the contact line, Kistler's dynamic contact angle model is implemented with a specified value of $\theta_a$ and $\theta_r$. In the absence of a slip boundary condition, the VOF solver utilizes the cell face normal velocities in computational cells immediately above the solid contact boundary to estimate $U_{CL}$ for use in the Kistler model and to advect volume fractions. Fluid properties are for a water-air system at approximately ambient temperature, see Table \ref{tab:1.1}. Initial drop diameters, $D$, range from 0.7 mm to 1.0 mm, meaning at least 70 cells per drop diameter. This has been shown to give convergence in other coalescence simulations \cite{Attarzadeh2017,Chu2018,Moghtadernejad2015,Graham2012}. Finally, gravity is neglected to facilitate comparison against prior computational literature on sessile drop coalescence. For model verification versus experiment, the domain size is amended to 2.6 mm x 2.6 mm x 2.6 mm with cubic cells of edge length $\epsilon = 20 \ \mu$m to accommodate $D$ values ranging from 1.5 mm to 2.0 mm. Also, the gravity vector is enforced in the negative $z$-axis direction with magnitude 9.807 m/s$^{2}$ to match the experimental conditions in our Earth-based experiments.

In the numerical solution scheme, the advection equation for $\alpha$ is first solved via the multi-dimensional universal limiter with explicit solution (MULES) method at a timestep, $\Delta t$, that satisfies a Courant number of $C \leq 0.2$ in each computational cell. For computations in three dimensions, the Courant number is defined as
\begin{equation}
  C = \frac{|u_x|\Delta t}{\epsilon}+\frac{|u_y|\Delta t}{\epsilon}+\frac{|u_z|\Delta t}{\epsilon}
\end{equation}
and governs how rapidly data is translated in the simulation domain to ensure numerical stability. Next, to maintain the contact angle prescribed by the Kistler model, the interface orientation obtained from the advection equation is corrected at each time step. Then, the liquid-gas interface is smoothed before calculation of the interfacial curvature $\kappa$. Finally, the pressure implicit with splitting of operators (PISO) algorithm \cite{Issa1986} is employed to solve the momentum and continuity equations. Output files containing data from the computations are generated at user-specified time intervals for subsequent post-processing.

\section{Characteristic Scales and Non-Dimensional Numbers}
For the inertial coalescence of capillary drops, motions respond on the inertial-capillary timescale $\tau=(\rho_{l}R^{3}/\sigma_{lg})^{1/2}$ where, for two identical coalescing drops, $R$ is the initial radii of the drops. This allows the dynamics to be described in terms of the non-dimensional timescale $t^*=t/\tau$. The characteristic length scale is taken as the initial diameter of a drop, $D=2R$. Capillary drops are distinguished by Bond numbers $Bo=(\rho_l-\rho_g)|\bm{g}|D^2/\sigma_{lg}$ less than unity for which surface tension dominates over gravity. Other relevant non-dimensional numbers include the Reynolds number $Re=\rho_lu_{ch}D/\mu_l$, Ohnesorge number $Oh=\mu_l/(\rho_l\sigma_{lg}D)^{1/2}$, and Capillary number $Ca=\mu_lu_{ch}/\sigma_{lg}$ where $u_{ch}$ is a characteristic velocity in the liquid phase. In this work, $u_{ch}$ is specified as the maximum value of the mass-averaged velocity in the liquid phase over the course of a coalescence event,
\begin{equation}
  u_{ch} \equiv max\Bigg(\frac{\int_{V_l}^{} \rho|\bm{u}|dV}{\int_{V_l} \rho \, dV}(t)\Biggr),
\end{equation}
where $V_l$ is the total liquid phase volume and $dV$ is the volume of a mesh element.

\section{Sweeping Metrics}
From a top view of the coalescing drops, a projected swept area $\mathscr{P}_\infty$ is traced as the combining liquid drops progress to an equilibrium state, Figure \ref{fig:1.2}. The fractional increase in projected swept, relative to initial projected area $\mathscr{P}_0=2\pi R^2$, is
\begin{equation}
  \mathscr{P}^* \equiv \frac{\mathscr{P}_\infty}{\mathscr{P}_0}.
\end{equation}
Similarly, from a bottom view, a wetted area $\mathscr{W}_\infty$ is swept in time, Figure \ref{fig:1.2}. The fractional increase in wetted swept, relative to initial wetted area $\mathscr{W}_0=2\pi r^2$, is
\begin{equation}
  \mathscr{W}^* \equiv \frac{\mathscr{W}_\infty}{\mathscr{W}_0},
\end{equation}
with $r=R\sin{\theta_0}$ the base wetted radius of the initial drop. Alternatively, a net cumulative metric can be used, $\mathscr{P}^{**} \equiv \mathscr{P}_\infty/ \mathscr{P}_0 - 1$, $\mathscr{W}^{**} \equiv \mathscr{W}_\infty/\mathscr{W}_0 - 1$. Since the denominator $\mathscr{W}_0$ vanishes as $\theta_0 \rightarrow{180 \degree}$, we also use regularized forms, $\mathscr{W}^* \, (\mathscr{W}_0/\mathscr{P}_0)$ and $\mathscr{W}^{**} \, (\mathscr{W}_0/\mathscr{P}_0)$, for respective comparison to $\mathscr{P}^*$ and $\mathscr{P}^{**}$. Time progressions of the projected ($x_p$ \& $y_p$) and wetted ($x_w$ \& $y_w$) extensions in the $x$-and $y$-axes are also evaluated, Figure \ref{fig:1.2}. As discussed in the results, projected extensions are normalized by $2R$ while wetted extensions are normalized by $r+R$ with non-dimensional parameters indicated by an asterisk superscript ($^*$).

\section{Mechanical Energy Balance}
To interpret the coalescence sweeping results, we evaluate the mechanical energy balance of Dussan V. and Davis \cite{Dussan1986} using the VOF simulations. This continuum mechanical energy balance involves sharp interface integrals over volumes, surfaces, and contours. For these evaluations, the VOF contour $\alpha = 0.5$ is identified as the sharp interface location and the contact line is estimated as intersection of the drop surface and the substrate. This energy balance separates reversible energy changes from dissipative losses,
\begin{equation} \label{eq:mechengbal}
  \frac{d}{dt} \, \Big(\mathscr{A}+\mathscr{K}\Big)=-\mathscr{D}<0,
\end{equation}
where $\mathscr{A}=A_{sl}\sigma_{sl}+A_{lg}\sigma_{lg}+A_{sg}\sigma_{sg}$ is interfacial energy, $\mathscr{K} = \int_{V}^{} \frac{1}{2} \, \rho \, (\bm{u}\cdot\bm{u}) \, dV$ the kinetic energy, and $\mathscr{D}$ the dissipation rate. The interfacial energy is defined as the sum of the solid-liquid ($sl$), liquid-gas ($lg$), and solid-gas ($sg$) interfacial areas, $A$, weighted by respective surface energies, $\sigma$. The dissipation rate $\mathscr{D}=\mathscr{D}_{\theta} + \mathscr{D}_{\mu} + \mathscr{D}_{\beta}$ has possible contributions from contact line motion, $\mathscr{D}_{\theta}$, slip at the liquid-solid surface, $\mathscr{D}_{\beta}$, and from volumetric viscous effects, $\mathscr{D}_{\mu}$. The work done at the contact line by the uncompensated Young's force is fully dissipated enabling an evaluation, $\mathscr{D}_\theta = \int_{\gamma}^{} \sigma_{lg} (\cos{\theta_e} - \cos{\theta}) U_{CL} \, d\gamma$, where $\gamma$ is the contact line arc length and $\theta_e$ is the equilibrium contact angle from the Young-Dupr\'e equation, $\sigma_{sg}=\sigma_{sl}+\sigma_{lg}\cos{\theta_e}$. Regarding slip dissipation, $\mathscr{D}_{\beta} = 0$ is consistent with the no-slip condition $\beta = 0$ implemented in the VOF simulations. Additionally, taking $\beta = 0$ maintains a simulation scheme with no adjustable parameters. The viscous dissipation, $\mathscr{D}_\mu = \int_{V}^{} \mu \, \Phi \, dV$, is calculated in the standard way from the viscous dissipation function,
\begin{dmath}
  \Phi = 2\bigg[\Big(\frac{\partial u_x}{\partial x}\Big)^2+\Big(\frac{\partial u_y}{\partial y}\Big)^2+\Big(\frac{\partial u_z}{\partial z}\Big)^2\bigg]+\bigg(\frac{\partial u_y}{\partial x}+\frac{\partial u_x}{\partial y}\bigg)^2+\bigg(\frac{\partial u_z}{\partial y}+\frac{\partial u_y}{\partial z}\bigg)^2+\bigg(\frac{\partial u_x}{\partial z}+\frac{\partial u_z}{\partial x}\bigg)^2- \, \frac{2}{3}(\bm{\nabla}\cdot \, \bm{u})^2.
\end{dmath}
The volume $V$ in the integrals of $\mathscr{D}_\mu$ and $\mathscr{K}$ is taken as the combined volume $V$ of liquid and gas phases; that is, the full computational domain. 
Integration of Equation (\ref{eq:mechengbal}) in time yields that the change in kinetic plus interfacial energy must balance the dissipative losses over that time interval. Introducing the notation $\Delta\mathscr{A}(t) \equiv \mathscr{A}(t)-\mathscr{A}(0)$ and recognizing $\mathscr{K}(0)=0$ leads to
\begin{equation} \label{eq:intmechengbal}
  \Delta\mathscr{A}(t)+\mathscr{K}(t)=-\int_{0}^{t} \mathscr{D} \, dt \equiv - \overline{\mathscr{D}}(t).
\end{equation}
During a coalescence event, interfacial and kinetic energies convert back and forth while continuously being discounted by dissipation. Using Equation (\ref{eq:intmechengbal}), we track how this energy interchange occurs during sweeping. All energies are made non-dimensional via scaling by $\sigma_{lg}R^2$, with dimensionless parameters indicated by an asterisk ($^*$) as mentioned previously. As time progresses, the drop either jumps, at $t_{jump}$, or the kinetic energy transient dies out, at $t_{\infty}$, where, $t_{\infty}$ is defined as when oscillations in the contact line and liquid-gas interface have decayed to within $1\%$ of some long-time average position. At long enough times, one expects the loss of interfacial energy to balance dissipation,
\begin{equation}\label{eq:longtimeintmechengbal}
  \Delta\mathscr{A}(t_{\infty}) = - \overline{\mathscr{D}}(t_{\infty}).
\end{equation}
The difference between right and left hand sides of Equation (\ref{eq:intmechengbal}) serves as an additional assessment of the simulation quality. For all cases (Table~\ref{tab:1.3}), with Equation (\ref{eq:intmechengbal}) evaluated at $t_{jump}$ (superhydrophobic cases) or at $t_{\infty}$ (hydrophobic cases), the left- and right-hand sides of Equation (\ref{eq:intmechengbal}) agree to $15\%$ or better. This discrepancy may have various contributions: a) early phase error, since a small initial discrepancy in phase can lead to large discrepancies longer term for underdamped oscillators; b) neglected dissipation, since there may be effective slip near the contact line in VOF while $\beta=0$ was assumed in Equation (\ref{eq:intmechengbal}); and c) sharp interface location, since for the energy-dissipation balance, $\alpha = 0.5$ may not be appropriate as it is for mass conservation. Of course, a) and b) may be related since phase differences are well-known to be sensitive to damping. However, trends reported below based on the evaluation of Equation (\ref{eq:intmechengbal}), for example in the discussion of Figures~\ref{fig:1.8} and \ref{fig:1.9}, are not influenced by this discrepancy. Further examination of these matters is beyond our scope.

\section{Results and Discussion}

\subsection{Model Validation}
\begin{figure}[!t]
\centering
\includegraphics{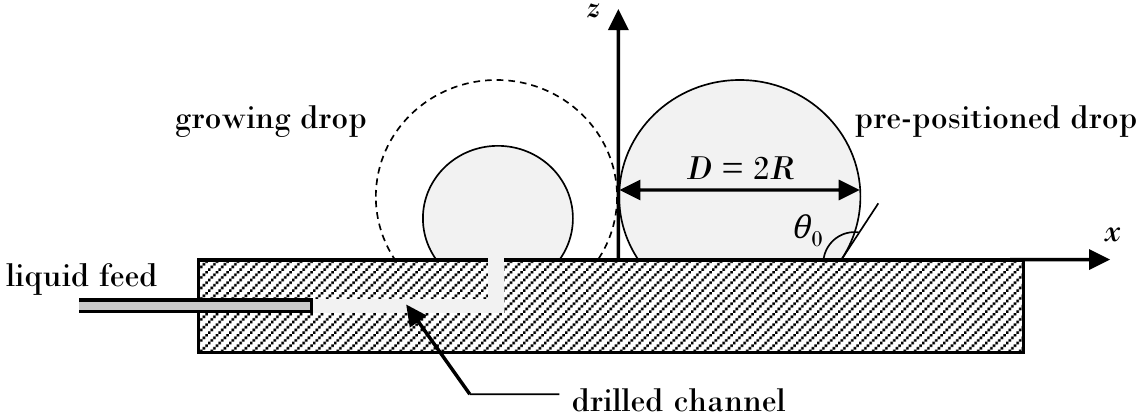}
\caption[]{Experimental setup (side view): liquid fed drop (left) grows until it touches a pre-positioned drop (right), initiating the coalescence event.}
\label{fig:1.4}
\end{figure}

\begin{figure}[t]
\resizebox{1.0\hsize}{!}{\includegraphics*{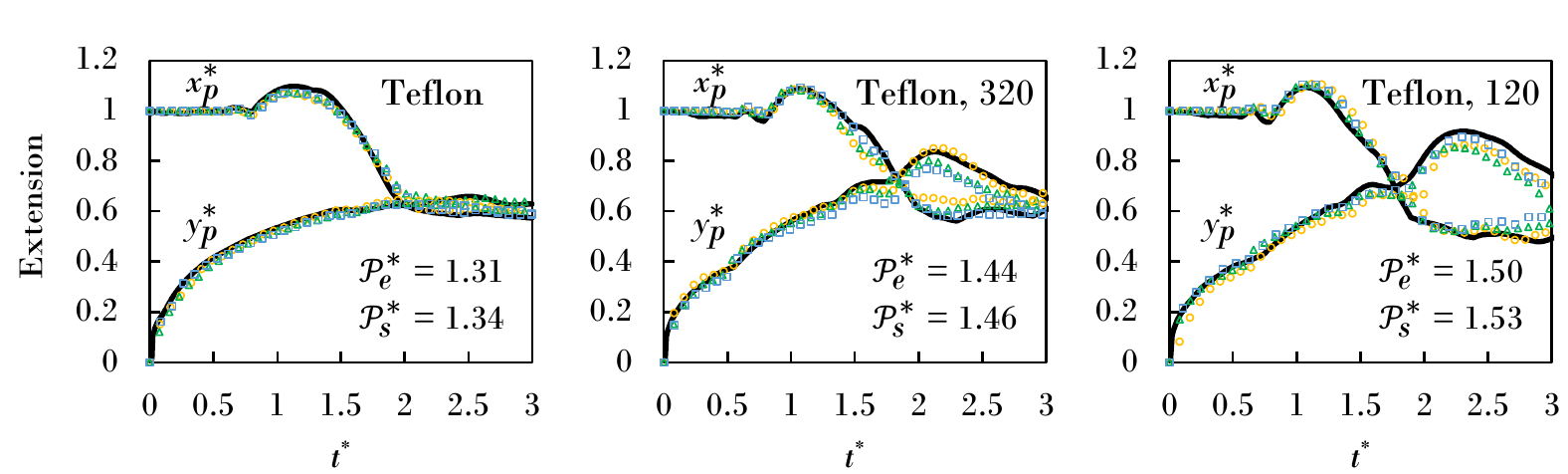}}
\caption[]{Validation: simulation (solid line) and experiment (markers) for three runs each (triangle, circle, and square markers) and three surfaces (Table \ref{tab:1.2}). Projected swept areas, for experiment $\mathscr{P}_e^*$ and simulation $\mathscr{P}_s^*$, as listed. No fitting parameters.}
\label{fig:1.5}
\end{figure}

The numerical model is verified against binary sessile drop coalescence experiments on various surfaces, with three coalescence trials performed on each substrate. A summary of the tested surfaces with their wettability metrics is found in Table \ref{tab:1.2}, where contact angle hysteresis is defined as $\Delta\theta=\theta_{a}-\theta_{r}$. 
\begin{table}[!b]
\centering
\caption[]{Experimental surfaces (chemistry, roughness) and wettability by water: initial contact angle $\theta_0$, advancing $\theta_a$, receding $\theta_r$, and contact angle hysteresis $\Delta \theta$.}
\begin{tabular}{|c||c|c|c|c|}
    \hline
    Surface & $\theta_{0}$ (\degree) & $\theta_a$ (\degree) & $\theta_r$ (\degree) & $\Delta\theta$ (\degree)\\
    \hline
    \hline
    Teflon & 110 & 120 & 75 & 45\\
    \hline
    Sanded Teflon, 320 grit & 141 & 154 & 76 & 78\\
    \hline
    Sanded Teflon, 120 grit & 151 & 159 & 102 & 57 \\
    \hline
\end{tabular}
\label{tab:1.2}
\end{table}
All sanded surfaces are prepared using silicon carbide sandpaper (Starcke Abrasives). For the experiments, coalescence is induced by slowly growing a drop via a 0.5 mm diameter through-hole in the surface until it touches a second pre-positioned drop (Figure \ref{fig:1.4}). An automatic dispensing system (ram\'e-hart part no. 100-22) is used to control the drop growth rate at 0.2 $\mu$L/s such that the drops have a minimal approach velocity. In all experiments, the drops are of approximately equal size at the time of coalescence (diameters within 5\% of each other) with $D$ in the range of 1.5 mm to 2.0 mm ($Bo\sim$ 0.3 to 0.5). Top and side view video cameras (Redlake MotionPro HS-3 and Redlake MotionXtra HG-XL, respectively) simultaneously record the coalescence dynamics at a frame rate of at least 3,500 frames per second. The side view gives information about the initial drop sizes and contact angles, while the top view ensures the drops are initially axisymmetric and gives the projected swept area and extensions. Comparing simulation against experiment, the time evolutions of $x_p^*$ and $y_p^*$ are in excellent agreement in all cases, Figure \ref{fig:1.5}, where data is plotted against the non-dimensional time. 
The non-dimensional projected swept areas resulting from the coalescence dynamics are also found to be comparable in all cases, with the simulation thought to slightly exceed the experiment due to the exact matching of the initial drop sizes. For a comparison of fluid dynamics by top view between simulation and experiment for select cases, see Video S1 of the Supplementary Material. Note that Wasserfall and collaborators \cite{Wasserfall2017} have independently verified the modified interFoam solver by simulating free drop oscillations at various Reynolds numbers. 

\subsection{Parametric Study}
After validating the numerical model, a parametric study of the surface wettability influence on sweeping is undertaken. Table \ref{tab:1.3} provides the various contact angles tested over the range $90 \degree \le \theta_0 \le 180 \degree$. In all cases, except for $\theta_0=180\degree$, the contact angle hysteresis was set to $\Delta\theta=15\degree$ with $\theta_a=\theta_0+5\degree$. Table \ref{tab:1.3} also provides the relevant dimensionless numbers associated with the dynamics of the coalescence event. For all simulations, the Reynolds number confirms laminar flow while the Ohnesorge and Capillary numbers indicate that capillary and inertia forces dominate over viscous effects. Additionally, recall that gravity is set to zero in these studies so $Bo=0$. For each simulation in the hydrophobic regime ($90\degree\le\theta_0<150\degree$), the left- and right-hand sides of Equation (\ref{eq:intmechengbal}) agree within 15\% of each other as $t\rightarrow{} t_{\infty}$. This indicates that the interfacial energy release is well-balanced by dissipation at long times when the kinetic energy approaches zero. For all superhydrophobic drop coalescence cases ($\theta_0 \ge 150\degree$), the liquid jumps away from the surface during the coalescence event. For these, Equation (\ref{eq:intmechengbal}) is evaluated at the time of drop jumping ($t_{jump}$). The agreement between the left- and right-hand sides falls within a few percent of each other. These energy analyses provide an additional indication of the simulation fidelity.
\begin{table}[!b]
\centering
\caption[]{Simulation parameters: surface wettability angles $\theta_0$, $\theta_a$, $\theta_r$, and $\Delta \theta$; initial diameter $D$; characteristic velocity $u_{ch}$; Reynolds $Re$, Ohnesorge $Oh$, and Capillary $Ca$ numbers.}
\begin{tabular}{|c||c|c|c|c||c|c||c|c|c|}
    \hline
    No. & $\theta_0$ (\degree) & $\theta_a$ (\degree) & $\theta_r$ (\degree) & $\Delta\theta$ (\degree) & $D$ (mm) & $u_{ch}$ (m/s) & $Re$ & $Oh$ & $Ca$\\
    \hline
    \hline
    1 & 90 & 95 & 80 & 15 & 1.00 & 0.21 & 205 & 0.004 & 0.003\\
    \hline
    2 & 101 & 106 & 91 & 15 & 0.88 & 0.23 & 203 & 0.004 & 0.003\\
    \hline
    3 & 113 & 118 & 103 & 15 & 0.80 & 0.25 & 203 & 0.004 & 0.004\\
    \hline
    4 & 123 & 128 & 113 & 15 & 0.76 & 0.27 & 203 & 0.004 & 0.004\\
    \hline
    5 & 131 & 136 & 121 & 15 & 0.74 & 0.27 & 203 & 0.004 & 0.004\\
    \hline
    6 & 141 & 146 & 131 & 15 & 0.72 & 0.28 & 201 & 0.004 & 0.004\\
    \hline
    7 & 150 & 155 & 140 & 15 & 0.71 & 0.30 & 214 & 0.004 & 0.004\\
    \hline
    8 & 160 & 165 & 150 & 15 & 0.71 & 0.31 & 223 & 0.004 & 0.004\\
    \hline
    9 & 170 & 175 & 160 & 15 & 0.71 & 0.32 & 224 & 0.004 & 0.004\\
    \hline
    10 & 180 & 180 & 180 & 0 & 1.00 & 0.27 & 272 & 0.004 & 0.004\\
    \hline
\end{tabular}
\label{tab:1.3}
\end{table}

\begin{figure}[t]
\centering
\includegraphics{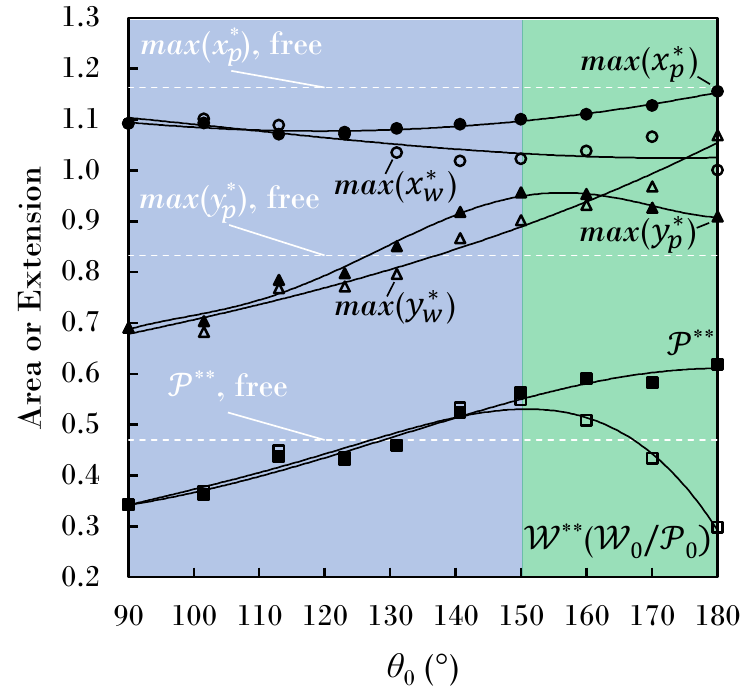}
\caption[]{Swept areas (squares) and maximum extensions ($x$, circles; $y$, triangles), as influenced by wettability $\theta_0$, with guide (solid fit lines). Binary free drop coalescence case (white dashed). Regimes: hydrophobic (blue) and superhydrophobic (green).}
\label{fig:1.6}
\end{figure}

\begin{figure}[t]
\resizebox{1.0\hsize}{!}{\includegraphics*{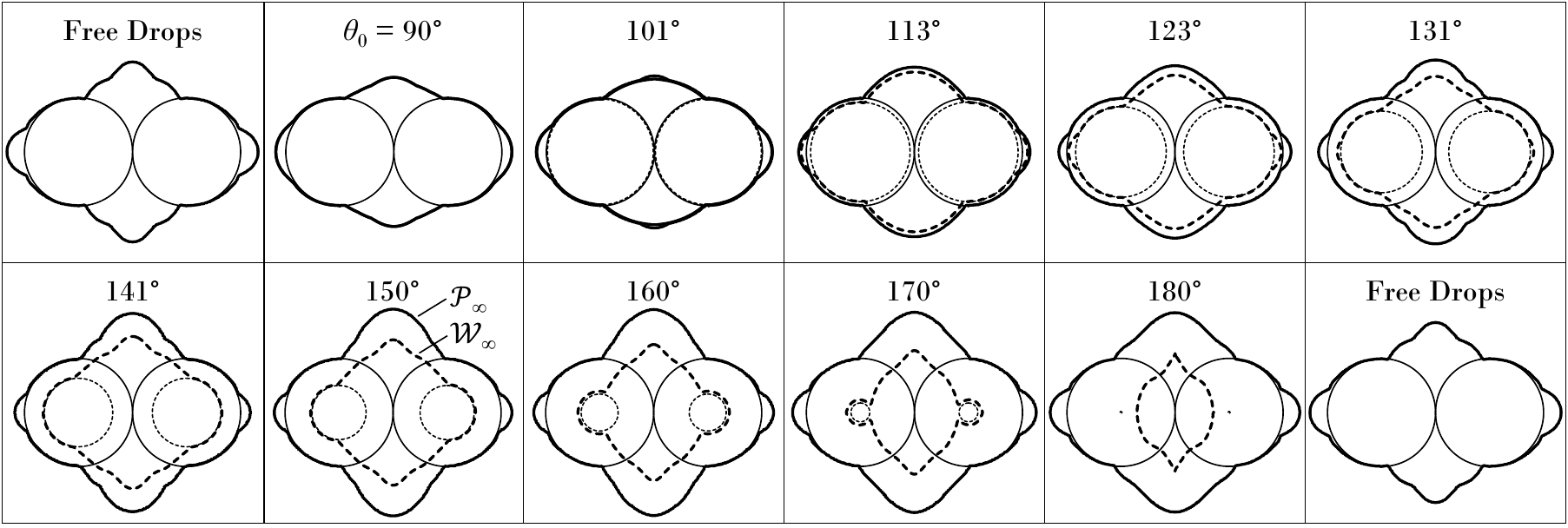}}
\caption[]{Cumulative swept areas: wetted $\mathscr{W}_\infty$ (thick dashed) and projected $\mathscr{P}_\infty$ (thick solid), with initial wetted (thin dashed) and initial projected (thin solid) areas, for reference. All areas are scaled, for comparison.}
\label{fig:1.7}
\end{figure}

Sweeping and extension results for both the hydrophobic and superhydrophobic regimes are plotted in Figure \ref{fig:1.6}, along with the free drop coalescence case (white dashed lines). 
In the hydrophobic regime (blue shading), net swept areas $\mathscr{P}^{**}$ and $\mathscr{W}^{**}(\mathscr{W}_0/\mathscr{P}_0)$ are observed to increase with $\theta_0$ which manifests from broader liquid extensions in the $y$-axis during the coalescence event. Additionally, when $\theta_0$ increases, the drop footprints become smaller and more separated which benefits $\mathscr{W}^{**}(\mathscr{W}_0/\mathscr{P}_0)$ because the surface area between the drop footprints is always traversed by liquid. The maximum extent of liquid deformation in the $x$-axis is observed to be similar in all cases and results from capillary wave propagation after coalescence initiation. Furthermore, in the hydrophobic regime, $\mathscr{P}^{**}$ and $\mathscr{W}^{**}(\mathscr{W}_0/\mathscr{P}_0)$ are nearly identical indicating that a similar amount of net wetted and projected area is swept by the coalescence event. In the superhydrophobic regime (green shading), $\mathscr{P}^{**}$ continues to increase with $\theta_0$ while $\mathscr{W}^{**}(\mathscr{W}_0/\mathscr{P}_0)$ trends downward as $\theta_0 \rightarrow \, 180\degree$. For $\mathscr{W}^{**}(\mathscr{W}_0/\mathscr{P}_0)$, the enhanced hydrophobicity compels less liquid contact with the surface which reduces wetted contact sweeping. Additionally, as $\theta_0$ increases, the liquid departs from the surface at earlier times which cuts short further wetted contact sweeping. Figure \ref{fig:1.7} illustrates the cumulative wetted and projected swept areas at various $\theta_0$, as normalized by the initial diameter of the drops. Initial drop projections and wetted contacts are also included for reference.

\begin{figure}[t]
\resizebox{1.0\hsize}{!}{\includegraphics*{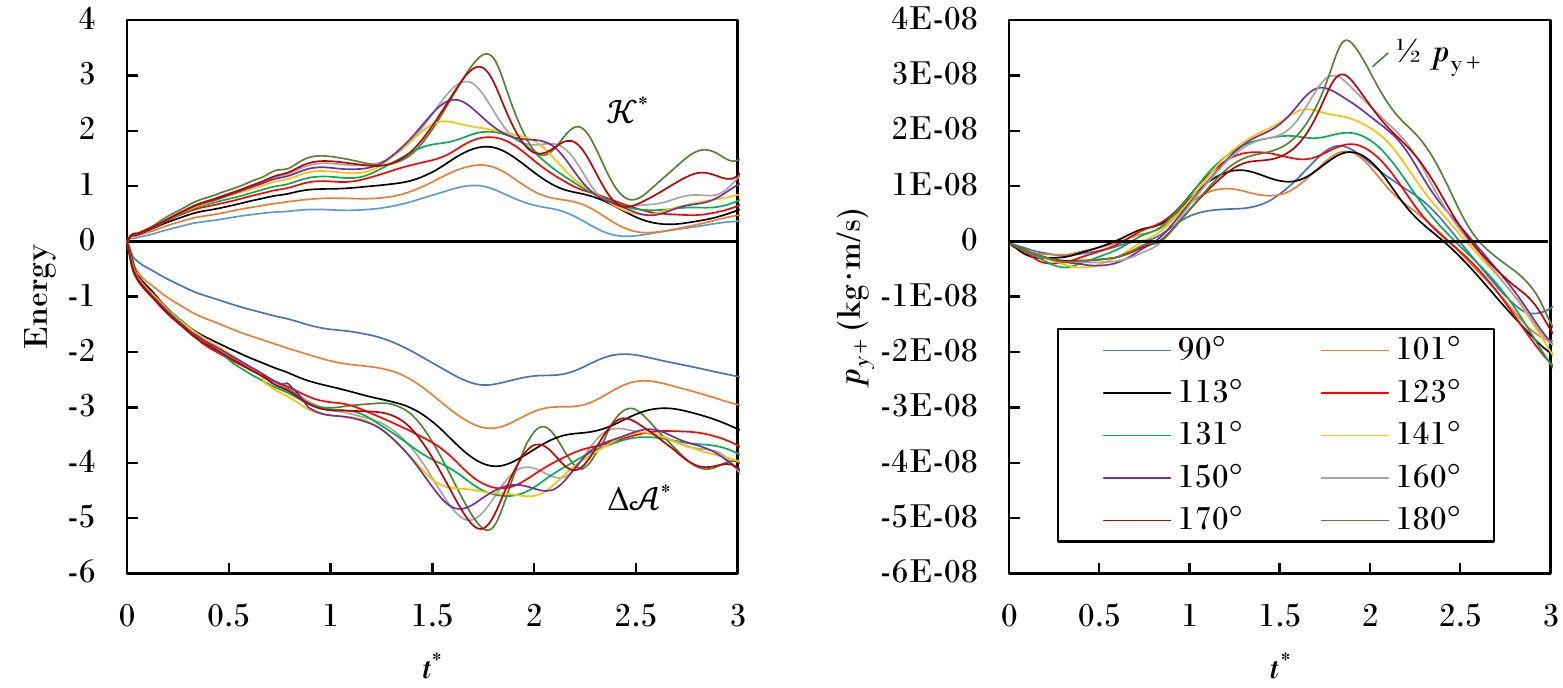}}
\caption[]{Evolution (left) of kinetic energy $\mathscr{K}^*$ and change in interfacial energy $\Delta\mathscr{A}^*$. Evolution (right) of total plane-normal momentum $p_{y+}$ ($y$-component in positive $y$-direction).}
\label{fig:1.8}
\end{figure}

To further elucidate the swept area results, the transients in interfacial, kinetic, and dissipative energies are analyzed. Figure \ref{fig:1.8} plots $\Delta\mathscr{A}^*$ and $\mathscr{K}^*$ versus $t^*$, where it is observed that the release of interfacial energy in time correlates directly with the evolution of the kinetic energy, as expected. 
\begin{figure}
\resizebox{1.0\hsize}{!}{\includegraphics*{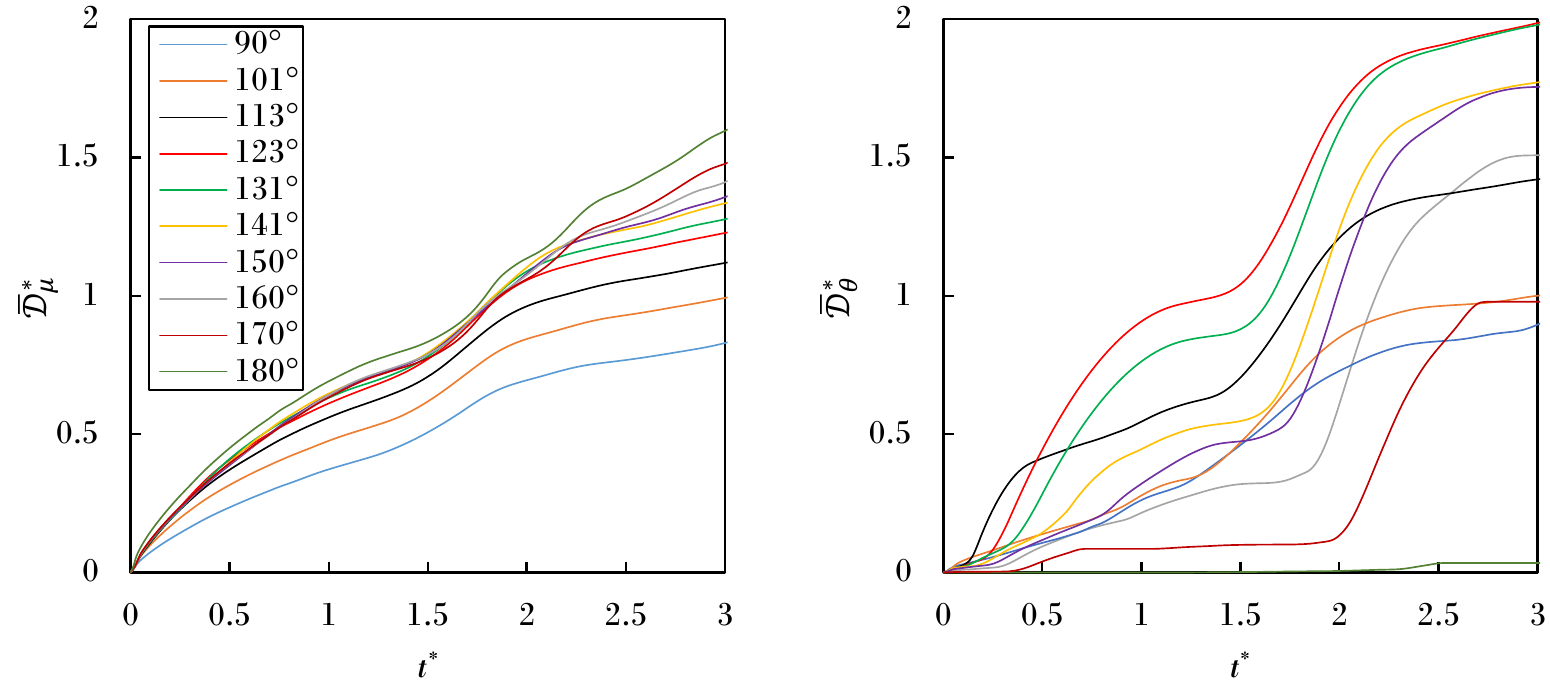}}
\caption[]{Evolution of cumulative energy losses: bulk viscous $\overline{\mathscr{D}}_{\mu}^*$ (left) and contact line $\overline{\mathscr{D}}_{\theta}^*$ (right) contributions to the overall dissipation $\overline{\mathscr{D}}^* = \overline{\mathscr{D}}_{\mu}^* + \overline{\mathscr{D}}_{\theta}^*$.}
\label{fig:1.9}
\end{figure}
In general, the kinetic energy transients are greater in magnitude at larger $\theta_0$ values. Additionally, it is noted that $max(\mathscr{K}^*)$ always occurs during the initial expansion of the liquid in the $y$-direction, specifically after the maximum $x$-axis extension is reached and the capillary waves are reflected back towards the coalescence initiation point ($1.5 < t^* < 2.0$). As $max(\mathscr{K}^*)$ increases with $\theta_0$ during this time, the liquid obtains a correspondingly larger $y$-component of momentum in the positive (and negative) $y$-axis domain, Figure \ref{fig:1.8}. This eventually leads to greater $y$-axis maximum extensions during $2.0 < t^* < 2.5$ which increases the amount of area swept. In terms of energy loss during coalescence, Figure \ref{fig:1.9}, the dimensionless bulk viscous dissipation, $\overline{\mathscr{D}}_{\mu}^* \equiv \int_0^t \mathscr{D}_\mu\,dt /\sigma_{lg}R^2$, is monotonically increasing in $\theta_0$ nearly uniformly in $t^*$ over the range while contact line dissipation, $\overline{\mathscr{D}}_{\theta}^* \equiv \int_0^t \mathscr{D}_\theta\,dt /\sigma_{lg}R^2$, is rather mixed. For example, at $t^*=3$ the order of increasing $\overline{\mathscr{D}}_{\theta}^*$ is $\theta_0 = \{180\degree, 90\degree, 170\degree, 101\degree, 113\degree, 160\degree, 150\degree, 141\degree, 131\degree, 123\degree \}$. The maximum in $\overline{\mathscr{D}}_{\theta}^*$ that occurs near $123 \degree$ is evidently the result of competition between lessening perimeter and increasing vigor of contact line motion owing to heightening center of mass. Decreased total dissipation, of course, allows for a larger conversion of interfacial energy to kinetic energy.

A comparison of the projected swept areas for sessile drop coalescence is also made with the free drop coalescence case, Figures 6 and 7. For the pairwise coalescence of two free drops, $\mathscr{P}^{**}$ = 0.44 with $max(x_p^*)$ = 1.16 and $max(y_p^*)$ = 0.83. The value of $max(x_p^*)$ is slightly greater than all the sessile drop cases because the capillary wave traveling on the bottom side of the liquid is undisturbed by a surface. This allows it to eventually reach the extreme edge of the drop which amplifies the final liquid extension. In contrast, the value of $max(y_p^*)$ is surpassed by the sessile drop case at $\theta_0\sim130\degree$. This is a result of the surface breaking the symmetry of the coalescence dynamics. As $\theta_0$ trends towards 180\degree, the drops become more spherical similar to the free drop case. However, since the surface cannot be penetrated by fluid, when the liquid bridge impinges on the substrate the fluid is redirected upwards and can lead to enhanced sweeping. For $\theta_0>130\degree$, the larger values of $max(y_p^*)$ achieved versus the free drop case allow sessile drops to attain greater $\mathscr{P}^{**}$ values.

\begin{figure}[!t]
\resizebox{1.0\hsize}{!}{\includegraphics*{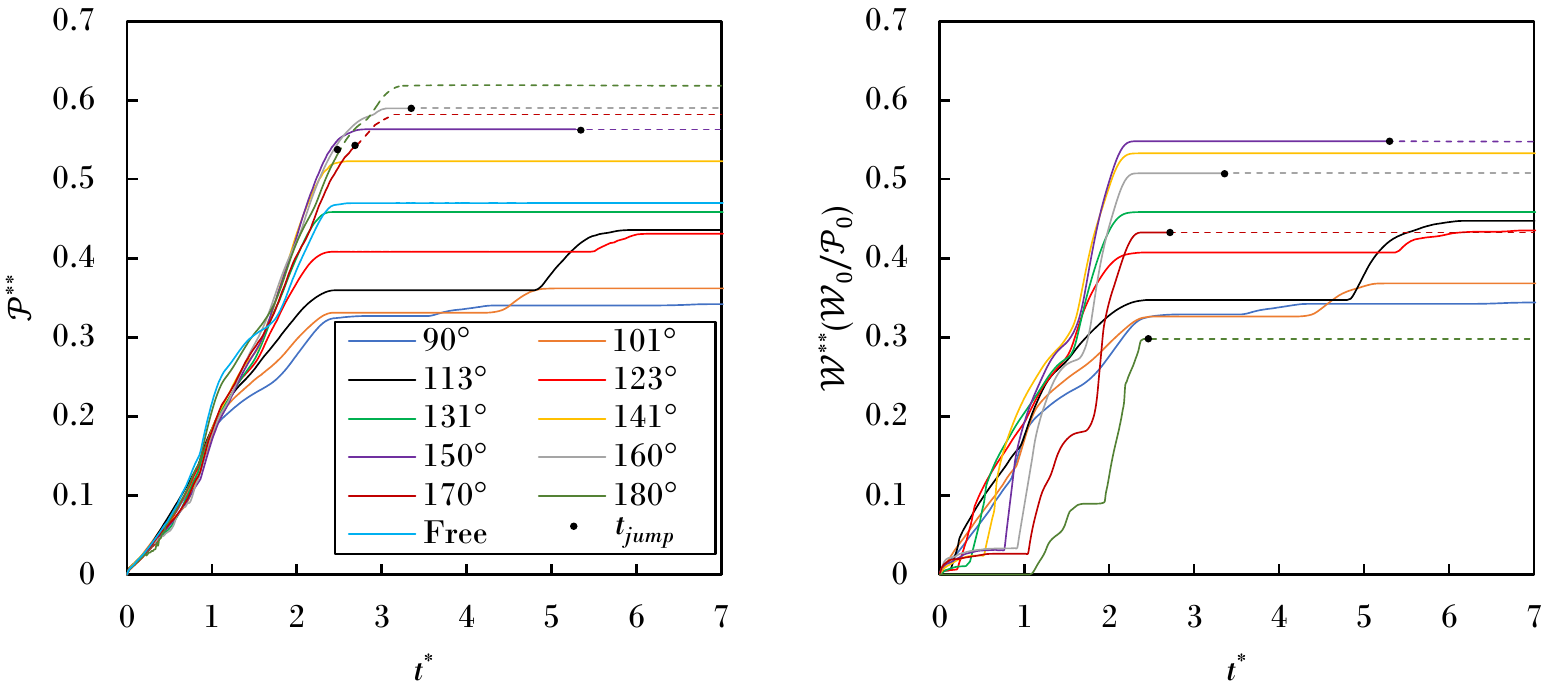}}
\caption[]{Evolution of net swept areas: projected $\mathscr{P}^{**}$ (left) and wetted $\mathscr{W}^{**}(\mathscr{W}_0/\mathscr{P}_0)$ (right). Solid circles indicate drop jumping, at which wetted trajectories terminate (extensions, dashed).} 
\label{fig:1.10}
\end{figure}

The time duration of sweeping is additionally investigated with Figure \ref{fig:1.10} showing $\mathscr{P}^{**}$ (left) and $\mathscr{W}^{**}(\mathscr{W}_0/\mathscr{P}_0)$ (right) versus $t^*$ for various $\theta_0$ values. It is observed that the evolution of $\mathscr{P}^{**}$ is nearly the same through $t^*$ = 1.125 for all sessile drop coalescence cases, except $\theta_0=180\degree$ which evolves similarly to the free drop case. This time frame corresponds to early liquid bridge growth and the initial capillary wave propagation to $max(x_p^*)$. Deviations occur after this time, but in all cases more than 80\% of $\mathscr{P}^{**}$ is swept by the time $max(y_p^*)$ is reached at $t^*\sim$ 2.5. 
This indicates that a majority of the sweeping happens rapidly via the attainment of $max(x_p^*)$ and then $max(y_p^*)$ in succession during the first full oscillation of the liquid. For $\theta_0>130\degree$, the remaining area to be swept is achieved during liquid retraction immediately after $max(y_p^*)$ is attained, similar to the free drop coalescence case. Note that in the superhydrophobic regime ($\theta_0\ge150\degree$), the projected swept area can continue to evolve after the instant of drop jumping; see the dashed lines in Figure \ref{fig:1.10}. For $\theta_0<130\degree$, $\mathscr{P}^{**}$ plateaus after $t^*\sim$ 2.5 before eventually reaching its final value as the liquid settles to an equilibrium shape. The time span before this settling is observed to increase with $\theta_0$ due to the increased kinetic energy in the liquid. For $\mathscr{W}^{**}(\mathscr{W}_0/\mathscr{P}_0)$, analogous sweeping behavior is observed in time. Although the progression of $\mathscr{W}^{**}(\mathscr{W}_0/\mathscr{P}_0)$ varies with differing wettabilities early on due to the different times of liquid bridge impingement with the surface, more than $\sim 80\%$ of $\mathscr{W}^{**}(\mathscr{W}_0/\mathscr{P}_0)$ is swept by $t^*\sim 2.5$ in all cases. See the Supplementary Material for select videos illustrating the sweeping behavior in time, Videos S2 and S3.

We end this section with a summary of the significance of these results to engineering applications. Heat transfer coefficients for dropwise condensation are an order-of-magnitude greater than those for filmwise condensation, as is well known \cite{Cho2016}. Dropwise condensation on hydrophobic surfaces can be maintained by effective coalescence sweeping strategies. Sweeping of wetted area removes recently nucleated drops, while sweeping of projected area induces coalescence of larger-scale drops. Eventually, drops grow large enough via coalescence to be detached via gravity-driven sliding which results in additional sweeping of surface area. These mechanisms act to delay or avoid filmwise condensation. 

In addition to use in dropwise condensers, hydrophobic surfaces are attractive for self-cleaning surfaces, anti-frost coatings, and water harvesters, among others. On self-cleaning surfaces, enhanced hydrophobicity may assist with the collection of contaminants and other foreign debris by increasing the amount of area swept by the liquid phase. Similarly, for anti-frost coatings, a greater extent of sweeping via increased surface hydrophobicity may improve the removal efficiency of drops before freezing can occur. For water harvesting devices, coalescence is responsible for growing drops to a critical size for departure and collection. Larger swept areas can enhance growth rates, as in dropwise condensation, and lead to more frequent drop departure and higher rates of collection.

\section{Concluding Remarks}
Sweeping by sessile drop coalescence is studied in an idealized setup. Coalescence between two identical drops, initially at rest on a flat and non-porous hydrophobic substrate, is initiated by arranging the drops to just touch. We first validate our simulations against experiment for three different material systems and then computationally study swept areas for increasing $\theta_0$ in the hydrophobic and superhydrophobic regimes. The systems studied are characterized by initial contact angles $90\degree \le \theta_0 \le 180\degree$ and contact angle hysteresis $\Delta \theta = 15\degree$. Our post-processing interpretation of the computations uses the hydrodynamic theory of moving contact lines.

\begin{itemize}
   \item
   Simulation modeling of the experiment presents a number of challenges including rapidly moving contact lines, inertially-dominated dynamics, and possible drop jumping. Using an empirically-based model of moving contact lines, our simulation reasonably captures the observed swept area evolutions for water on three different substrates, with no fitting parameters.
 \end{itemize} 
 \begin{itemize}
   \item
   In this work, coalescence is driven solely by excess interfacial energy as the event starts and ends at rest. Initially, interfacial energy is converted to kinetic energy and then back again while simultaneously being degraded by bulk viscous and contact line dissipation. Eventually, kinetic energy approaches zero and the interfacial energy release is expected to equal the total dissipation. We track energies and losses in time with the energy balance, Equation (\ref{eq:intmechengbal}), ultimately evaluated at $t_{\infty}$ (hydrophobic cases) or $t_{jump}$ (superhydrophobic cases). For all simulation events reported, the left- and right-hand sides of Equation (\ref{eq:intmechengbal}) agree to better than $15\%$.
 \end{itemize} 
 \begin{itemize}
   \item As wettability decreases ($\theta_0$ increases), wetted area initially tracks projected area up to about $\theta_0 = 150\degree$, at which angle they begin to deviate. Soon after, wetted area reaches a maximum and turns over owing to a smaller wetted perimeter and possibly early termination of the evolution by drop jumping. At jumping, wetted perimeter vanishes by shrinking to a point. Interestingly, the onset of jumping happens to coincide with the commonly accepted start of the superhydrophobic regime, $\theta_0 = 150\degree$. Our observation of coalescence-induced drop jumping in the superhydrophobic regime is consistent with other studies \cite{Farokhirad2015,Wang2017a}.
 \end{itemize}
 \begin{itemize}
   \item Viscosity is the only form of dissipation operative in free drop coalescence while sessile drop coalescence additionally has contact friction. Contact line dissipation can contribute more than half the total dissipation, according to our simulations, depending on wettability. It is perhaps surprising then that, by $\theta_0 = 130 \degree$, the sessile drop $\mathscr{P}^{**}$ can exceed the free drop $\mathscr{P}^{**}$ and by $\theta_0 = 180 \degree$, the sessile $\mathscr{P}^{**}$ exceeds the free $\mathscr{P}^{**}$ by more than $30\%$. One concludes that the increasing energy loss by contact line dissipation is outweighed by enhanced dynamical vigor owing to the higher center of mass (COM) that correlates with increasing $\theta_0$, or decreasing wettability.
 \end{itemize}
 \begin{itemize}
   \item The COM motion with and without substrate relates to jumping. For the free drop coalescence event, there is no net force in any direction exerted on the system. This implies, by Newton's law, that the system COM remains stationary throughout the event. With the introduction of the substrate, a vertical velocity component of the system COM motion becomes possible. That is, the liquid contact with the substrate introduces i) a pressure that acts vertically and ii) a contact line force around the perimeter due to surface tension, having a vertical component. As a consequence, the system COM can move upward even while its $xy$-projection remains stationary by no net horizontal force. In summary, jumping cannot occur obliquely and, to occur, must overcome the surface tension adhesion contribution at the contact line. This is increasingly likely to happen as wettability decreases ($\theta_0$ increases), consistent with our predictions.
 \end{itemize}
\noindent
In closing, it is important to note that coalescence infrequently takes place between two identical sessile drops in practical applications. As such, future studies that investigate the effects of drop size and/or contact angle mismatch on sweeping would be of value. Additionally, the sweeping behavior of three or more drops with or without chain-reaction coalescence remains unexplored. Open questions persist as to how the number of drops and their positioning affect coalescence sweeping. Finally, a parametric study of contact angle hysteresis $\Delta \theta$ would be of benefit to parse whether $\theta_0$ or $\Delta \theta$ is the dominant parameter in determining the extent of coalescence-induced sweeping.

\begin{acknowledgement}
The authors acknowledge the financial support of NSF Award \#1637960, which supports an International Space Station experiment as in Figure \ref{fig:1.4}, planned for 2020, and NASA Space Technology Research Fellowship \#80NSSC17K0144. The authors also thank Glenn Swan for hardware fabrication support.

\vspace*{10px} \noindent\textbf{Author Contribution Statement}
\\

\noindent
J.M.L. and P.H.S. conceived the initial idea of the research. J.M.L. carried out the experiments and simulations while P.H.S. guided the work. J.M.L. analyzed the data. J.M.L. and P.H.S. wrote the paper.
\\

\vspace*{10px} \noindent\textbf{Supplementary Material Available}
\\

\noindent
Figure S1: Time response comparison of $x_p^*$ for the modified and unmodified interFoam solvers versus experiment.

\vspace*{5px} \noindent Video S1: Comparison of the top view fluid dynamics for experiment versus simulation. (Left): Teflon-water-air system where $\theta_0=110\degree$, $\theta_a=120\degree$, and $\theta_r=75\degree$. (Right): Sanded Teflon (120 grit)-water-air system where $\theta_0=151\degree$, $\theta_a=159\degree$, and $\theta_r=102\degree$.

\vspace*{5px} \noindent Video S2: Evolution of the non-dimensional projected swept area in time for $\theta_0=101\degree$, $\theta_a=106\degree$, and $\theta_r=91\degree$.

\vspace*{5px} \noindent Video S3: Evolution of the non-dimensional projected swept area in time for $\theta_0=141\degree$, $\theta_a=146\degree$, and $\theta_r=131\degree$.
\end{acknowledgement}


\begin{thebibliography}{}
\bibitem{Takeda2002}
K. Takeda, A. Nakajima, Y. Murata, K. Hashimoto, T. Watanabe, Jpn. J. Appl. Phys. \textbf{41}, 287 (2002)

\bibitem{Tsai2009}
P. Tsai, S. Pacheco, C. Pirat, L. Lefferts, D. Lohse, Langmuir \textbf{25}, 12293 (2009)

\bibitem{Heydari2013}
G. Heydari, E. Thormann, M. J\"arn, E. Tyrode, P. M. Claesson, J. Phys. Chem. C \textbf{117}, 21752 (2013)

\bibitem{Dash2014}
S. Dash, S. V. Garimella, Phys. Rev. E \textbf{89}, 042402 (2014)

\bibitem{Tavakoli2015}
F. Tavakoli, H. P. Kavehpour, Langmuir \textbf{31}, 2120 (2015)

\bibitem{Menchaca-Rocha2001}
A. Menchaca-Rocha, A. Mart\'inez-D\'avalos, R. N\'{u}{\~n}ez, S. Popinet, S. Zaleski, Phys. Rev. E \textbf{63}, 046309 (2001)

\bibitem{Nilsson2011}
M. A. Nilsson, J. P. Rothstein, J. Colloid Interface Sci. \textbf{363}, 646 (2011)

\bibitem{Kavehpour2015}
H. P. Kavehpour, Annu. Rev. Fluid Mech. \textbf{47}, 245 (2015)

\bibitem{Wisdom2013}
K. M. Wisdom, J. A. Watson, X. Qu, F. Liu, G. S. Watson, C.-H. Chen, Proc. Natl. Acad. Sci. \textbf{110}, 7992 (2013)

\bibitem{Boreyko2013}
J. B. Boreyko, C. P. Collier, ACS Nano \textbf{7}, 1618 (2013)

\bibitem{Beysens2006}
D. Beysens, C. R. Phys. \textbf{7}, 1082 (2006)

\bibitem{Meakin1992}
P. Meakin, Rep. Prog. Phys. \textbf{55}, 157 (1992)

\bibitem{Macner2014}
A. M. Macner, S. Daniel, P. H. Steen, Langmuir \textbf{30}, 1788 (2014)

\bibitem{Kim2011}
S. Kim, K. J. Kim, J. Heat Transfer \textbf{133}, 081502 (2011)

\bibitem{Eggers1999}
J. Eggers, J. R. Lister, H. A. Stone, J. Fluid Mech. \textbf{401}, 293 (1999)

\bibitem{Duchemin2003}
L. Duchemin, J. Eggers, C. Josserand, J. Fluid Mech. \textbf{487}, 167 (2003)

\bibitem{Ristenpart2006}
W. D. Ristenpart, P. M. McCalla, R. V. Roy, H. A. Stone, Phys. Rev. Lett. \textbf{97}, 064501 (2006)

\bibitem{Sellier2009}
M. Sellier, E. Trelluyer, Biomicrofluidics \textbf{3}, 022412 (2009)

\bibitem{Lee2012}
M. W. Lee, D. K. Kang, S. S. Yoon, A. L. Yarin, Langmuir \textbf{28}, 3791 (2012)

\bibitem{Hernandez-Sanchez2012}
J. F. Hern{\'{a}}ndez-S{\'{a}}nchez, L. A. Lubbers, A. Eddi, J. H. Snoeijer, Phys. Rev. Lett. \textbf{109}, 184502 (2012)

\bibitem{Sui2013}
Y. Sui, M. Maglio, P. D. M. Spelt, D. Legendre, H. Ding, Phys. Fluids \textbf{25}, 101701 (2013)

\bibitem{Eddi2013}
A. Eddi, K. G. Winkels, J. H. Snoeijer, Phys. Rev. Lett. \textbf{111}, 144502 (2013)

\bibitem{Mitra2015}
S. Mitra, S. K. Mitra, Phys. Rev. E \textbf{92}, 033013 (2015)

\bibitem{Gokhale2004}
S. J. Gokhale, S. DasGupta, J. L. Plawsky, P. C. Wayner Jr., Phys. Rev. E \textbf{70}, 051610 (2004)

\bibitem{Kapur2007}
N. Kapur, P. H. Gaskell, Phys. Rev. E \textbf{75}, 056315 (2007)

\bibitem{Liao2008}
Q. Liao, X. Zhu, S. M. Xing, H. Wang, Exp. Therm. Fluid Sci. \textbf{32}, 1647 (2008)

\bibitem{Lai2010}
Y.-H. Lai, M.-H. Hsu, J.-T. Yang, Lab Chip \textbf{10}, 3149 (2010)

\bibitem{Wang2010}
H. Wang, Q. Liao, X. Zhu, J. Li, X. Tian, J. Supercond. Nov. Magn. \textbf{23}, 1165 (2010)

\bibitem{Zhu2017}
G. Zhu, H. Fan, H. Huang, F. Duan, RSC Adv. \textbf{7}, 23954 (2017)

\bibitem{Somwanshi2018}
P. M Somwanshi, K. Muralidhar, S. Khandekar, Phys. Fluids \textbf{30}, 092103 (2018)

\bibitem{Andrieu2002}
C. Andrieu, D. A. Beysens, V. S. Nikolayev, Y. Pomeau, J. Fluid Mech. \textbf{453}, 427 (2002)

\bibitem{Narhe2004}
R. Narhe, D. Beysens, V. S. Nikolayev, Langmuir \textbf{20}, 1213 (2004)

\bibitem{Narhe2005}
R. Narhe, D. Beysens, V. S. Nikolayev, Int. J. Thermophys. \textbf{26}, 1743 (2005)

\bibitem{Beysens2006a}
D. A. Beysens, R. D. Narhe, J. Phys. Chem. B \textbf{110}, 22133 (2006)

\bibitem{Narhe2008}
R. D. Narhe, D. A. Beysens, Y. Pomeau, Europhys. Lett. \textbf{81}, 46002 (2008)

\bibitem{Narhe2009}
R. D. Narhe, M. D. Khandkar, P. B. Shelke, A. V. Limaye, D. A. Beysens, Phys. Rev. E \textbf{80}, 031604 (2009)

\bibitem{Boreyko2009}
J. B. Boreyko, C.-H. Chen, Phys. Rev. Lett. \textbf{103}, 184501 (2009)

\bibitem{Boreyko2010}
J. B. Boreyko, C.-H. Chen, Phys. Fluids \textbf{22}, 091110 (2010)

\bibitem{Wang2011}
F.-C. Wang, F. Yang, Y.-P. Zhao, Appl. Phys. Lett. \textbf{98}, 053112 (2011)

\bibitem{Peng2013}
B. Peng, S. Wang, Z. Lan, W. Xu, R. Wen, X. Ma, Appl. Phys. Lett. \textbf{102}, 151601 (2013)

\bibitem{Nam2013a}
Y. Nam, H. Kim, S. Shin, Appl. Phys. Lett. \textbf{103}, 161601 (2013)

\bibitem{Liu2014}
X. Liu, P. Cheng, X. Quan, Int. J. Heat Mass Transf. \textbf{73}, 195 (2014)

\bibitem{Liu2014a}
F. Liu, G. Ghigliotti, J. J. Feng, C.-H. Chen, J. Fluid Mech. \textbf{752}, 39 (2014)

\bibitem{Enright2014}
R. Enright, N. Miljkovic, J. Sprittles, K. Nolan, R. Mitchell, E. N. Wang, ACS Nano \textbf{8}, 10352 (2014)

\bibitem{Liu2015a}
X. Liu, P. Cheng, Int. Commun. Heat Mass Transf. \textbf{64}, 7 (2015)

\bibitem{Nam2015}
Y. Nam, D. Seo, C. Lee, S. Shin, Soft Matter \textbf{11}, 154 (2015)

\bibitem{Farokhirad2015}
S. Farokhirad, J. F. Morris, T. Lee, Phys. Fluids \textbf{27}, 102102 (2015)

\bibitem{Wang2016}
K. Wang, Q. Liang, R. Jiang, Y. Zheng, Z. Lan, X. Ma, RSC Adv. \textbf{6}, 99314 (2016)

\bibitem{Cha2016}
H. Cha, J. M. Chun, J. Sotelo, N. Miljkovic, ACS Nano \textbf{10}, 8223 (2016)

\bibitem{Attarzadeh2017}
R. Attarzadeh, A. Dolatabadi, Phys. Fluids \textbf{29}, 012104 (2017)

\bibitem{Wang2017}
K. Wang, Q. Liang, R. Jiang, Y. Zheng, Z. Lan, X. Ma, Langmuir \textbf{33}, 6258 (2017)

\bibitem{Sheng2017}
Q. Sheng, J. Sun, W. Wang, H. S. Wang, C. G. Bailey, J. Appl. Phys. \textbf{122}, 245301 (2017)

\bibitem{Mouterde2017}
T. Mouterde, T.-V. Nguyen, H. Takahashi, C. Clanet, I. Shimoyama, D. Qu{\'{e}}r{\'{e}}, Phys. Rev. Fluids \textbf{2}, 112001(R) (2017)

\bibitem{Wang2017a}
K. Wang, R. Li, Q. Liang, R. Jiang, Y. Zheng, Z. Lan, X. Ma, Appl. Phys. Lett. \textbf{111}, 061603 (2017)

\bibitem{Wasserfall2017}
J. Wasserfall, P. Figueiredo, R. Kneer, W. Rohlfs, P. Pischke, Phys. Rev. Fluids \textbf{2}, 123601 (2017)

\bibitem{Zhang2017}
P. Zhang, Y. Maeda, F. Lv, Y. Takata, D. Orejon, ACS Appl. Mater. Interfaces \textbf{9}, 35391 (2017)

\bibitem{Shi2018}
Y. Shi, G. H. Tang, Comput. Math. with Appl. \textbf{75}, 1213 (2018)

\bibitem{Chu2018}
F. Chu, Z. Yuan, X. Zhang, X. Wu, Int. J. Heat Mass Transf. \textbf{121}, 315 (2018)

\bibitem{Gao2018}
S. Gao, Q. Liao, W. Liu, Z. Liu, J. Phys. Chem. Lett. \textbf{9}, 13 (2018)

\bibitem{Chen2018}
Y. Chen, Y. Lian, Phys. Fluids \textbf{30}, 112102 (2018)

\bibitem{Gao2018a}
S. Gao, Q. Liao, W. Liu, Z. Liu, J. Phys. Chem. C \textbf{122}, 20521 (2018)

\bibitem{Yuan2019a}
Z. Yuan, R. Wu, X. Wu, Int. J. Heat Mass Transf. \textbf{135}, 345 (2019)

\bibitem{Wang2010b}
H. Wang, X. Zhu, Q. Liao, P. C. Sui, J. Supercond. Nov. Magn. \textbf{23}, 1137 (2010)

\bibitem{Ahmadlouydarab2014}
M. Ahmadlouydarab, J. J. Feng, J. Fluid Mech. \textbf{746}, 214 (2014)

\bibitem{Moghtadernejad2015}
S. Moghtadernejad, M. Tembely, M. Jadidi, N. Esmail, A. Dolatabadi, Phys. Fluids \textbf{27}, 032106 (2015)

\bibitem{Seo2017}
D. Seo, S. Oh, S. Shin, Y. Nam, Int. J. Heat Mass Transf. \textbf{114}, 934 (2017)

\bibitem{Mirjalili2019}
S. Mirjalili, C. B. Ivey, A. Mani, Int. J. Multiph. Flow \textbf{116}, 221 (2019)

\bibitem{Dodds2012}
S. Dodds, M. S. Carvalho, S. Kumar, J. Fluid Mech. \textbf{707}, 521 (2012)

\bibitem{Huang2019}
C.-H. Huang, M. S. Carvalho, S. Kumar, Phys. Rev. Fluids \textbf{4}, 044005 (2019)

\bibitem{Shikhmurzaev2007}
Y. D. Shikhmurzaev, \textit{Capillary Flows with Forming Interfaces}, 1st edn. (Chapman and Hall/CRC, New York, 2007)

\bibitem{Liu2016}
C.-Y. Liu, E. Vandre, M. S. Carvalho, S. Kumar, J. Fluid Mech. \textbf{808}, 290 (2016)

\bibitem{Kistler1993}
S. F. Kistler, in \textit{Wettability}, edited by J. C. Berg (Marcel Dekker, New York, 1993)

\bibitem{Weller1998}
H. G. Weller, G. Tabor, H. Jasak, C. Fureby, Comput. Phys. \textbf{12}, 620 (1998)

\bibitem{OpenFOAM}
http://www.openfoam.com/.

\bibitem{Brackbill1992}
J. U. Brackbill, D. B. Kothe, C. Zemach, J. Comput. Phys. \textbf{100}, 335 (1992)

\bibitem{Sikalo2005}
\v{S}. \v{S}ikalo, H.-D. Wilhelm, I. V. Roisman, S. Jakirli{\'{c}}, C. Tropea, Phys. Fluids \textbf{17}, 062103 (2005)

\bibitem{Roisman2008}
I. V. Roisman, L. Opfer, C. Tropea, M. Raessi, J. Mostaghimi, S. Chandra, Colloids Surfaces A \textbf{322}, 183 (2008)

\bibitem{Saha2009}
A. A. Saha, S. K. Mitra, J. Colloid Interface Sci. \textbf{339}, 461 (2009)

\bibitem{Graham2012}
P. J. Graham, M. M. Farhangi, A. Dolatabadi, Phys. Fluids \textbf{24}, 112105 (2012)

\bibitem{Xu2018}
J. Xu, Y. Chen, J. Xie, Int. J. Heat Mass Transf. \textbf{116}, 951 (2018)

\bibitem{Hoffman1975}
R. L. Hoffman, J. Colloid Interface Sci. \textbf{50}, 228 (1975)

\bibitem{Issa1986}
R. I. Issa, J. Comput. Phys. \textbf{62}, 40 (1986)

\bibitem{Dussan1986}
E. B. Dussan V., S. H. Davis, J. Fluid Mech. \textbf{173}, 115 (1986)

\bibitem{Cho2016}
H. J. Cho, D. J. Preston, Y. Zhu, E. N. Wang, Nat. Rev. Mater. \textbf{2}, 16092 (2016)

\end{thebibliography}
\end{document}